\pdfoutput=1
\documentclass[11pt,letterpaper]{article}

\usepackage[utf8]{inputenc}
\usepackage[T1]{fontenc}
\usepackage{newtxtext,newtxmath}
\usepackage[letterpaper,margin=1in,footskip=0.45in]{geometry}
\usepackage{microtype}
\usepackage{setspace}
\usepackage{graphicx}
\usepackage{booktabs,array}
\usepackage[font=small,labelfont=bf,labelsep=period,justification=justified,singlelinecheck=false]{caption}
\usepackage{enumitem}
\usepackage{needspace}
\usepackage[authoryear,round,sort]{natbib}
\usepackage{authblk}
\usepackage{titlesec}
\usepackage{fancyhdr}
\usepackage{xcolor}
\definecolor{navy}{RGB}{20,45,100}
\definecolor{notegray}{RGB}{110,110,110}
\usepackage[colorlinks=true,linkcolor=navy,citecolor=navy,urlcolor=navy,hyperfootnotes=false,
            pdftitle={Voice-Based AI Feedback in Higher Education: A Field Study on Quality and Actionability},
            pdfauthor={Thilo Tamme, Leander Schilling, Anton Hantel},
            pdfkeywords={conversational AI, voice-based feedback, student evaluations, survey mode effects, CASA, dynamic capabilities, higher education}]{hyperref}

\titleformat{\section}{\large\bfseries}{\thesection}{0.8em}{}
\titleformat{\subsection}{\normalsize\bfseries}{\thesubsection}{0.8em}{}
\titleformat{\paragraph}[runin]{\normalsize\bfseries\itshape}{}{0pt}{}[.]
\titlespacing*{\section}{0pt}{2.2ex plus 1ex minus .2ex}{1.2ex plus .2ex}
\titlespacing*{\subsection}{0pt}{1.8ex plus .8ex minus .2ex}{0.8ex plus .2ex}
\titlespacing*{\paragraph}{0pt}{1.2ex plus .5ex}{0.6em}
\newcommand{\SMSnote}{This research proposal was accepted, following double-blind peer review, for presentation at the
  46th Annual Conference of the Strategic Management Society (SMS), Berlin, Germany, October 17--20, 2026.}
\newcommand{\SMSfooter}{Accepted after double-blind peer review for the 46th SMS Annual Conference, Berlin, October 17--20, 2026}

\makeatletter
\newcommand{\acceptancenote}[1]{%
  \begingroup
  \renewcommand\thefootnote{}%
  \long\def\@makefntext##1{\noindent\footnotesize ##1}%
  \footnotetext{#1}%
  \endgroup}
\makeatother

\fancypagestyle{plain}{\fancyhf{}\fancyfoot[R]{\small\thepage}}

\newcommand{\hyp}[2]{\begin{quote}\textbf{#1}\ #2\end{quote}}

\title{\vspace{-1.2em}{\normalsize\scshape Research Proposal}\\[0.9em]
  {\LARGE\bfseries Voice-Based AI Feedback in Higher Education:}\\[0.3em]
  {\Large\bfseries A Field Study on Quality and Actionability}}

\author[1]{Thilo Tamme}
\author[1]{Leander Schilling}
\author[2]{Anton Hantel}
\affil[1]{Technical University of Munich}
\affil[2]{Massachusetts Institute of Technology}
\date{February 2026}

\begin{document}

\maketitle
\thispagestyle{plain}
\acceptancenote{\SMSnote}

% ---------------------------------------------------------------------
\begin{abstract}
\noindent
Student feedback is a critical input for continuous improvement in higher education, yet traditional end-of-semester written surveys suffer from low response rates, superficial answers, satisficing, and limited actionability. Voice-based conversational AI agents can conduct structured interviews at scale, asking open-ended questions and probing follow-ups, and may capture richer qualitative feedback than written instruments. This research proposal asks what characteristics, depth, and strategic value feedback collected through voice-based conversational AI agents has in higher education, and whether such agents elicit detailed, honest, and actionable feedback that can support instructional improvement and institutional learning. Drawing on survey-mode-effects research, the Computers As Social Actors (CASA) paradigm, and dynamic capabilities theory, we develop three hypotheses on elaboration, social presence and disclosure, and actionability. A completed technical pilot at the Technical University of Munich established the feasibility of telephone-based AI interviews; the main field study is planned for the ``Advanced Leadership Communication'' course at MIT Sloan with approximately 60 students, using a convergent mixed-methods design and a four-dimensional coding framework for actionability (specificity, thematic richness, constructive orientation, contextual depth). The study extends CASA theory to evaluative educational settings, characterizes voice-based AI as a distinct feedback modality, and links micro-level feedback quality to institutional sensing capabilities.
\end{abstract}

\vspace{0.3em}

% ---------------------------------------------------------------------
\section{Introduction}

Student feedback represents a critical input for continuous improvement in higher education, enabling instructors to refine pedagogical approaches, adjust course content, and enhance learning outcomes. However, traditional feedback mechanisms---primarily end-of-semester written surveys---suffer from well-documented limitations: low response rates, superficial responses, satisficing behavior, and limited actionability \citep{dillman2006,wambsganss2020}. Students often provide brief, generic comments that offer little guidance for meaningful course improvement. As higher education institutions face mounting pressure to demonstrate teaching effectiveness and adapt rapidly to changing student needs, the quality and utility of feedback data have become strategic concerns.

Recent advances in artificial intelligence, large language models, and conversational AI offer promising alternatives to traditional survey methods. Voice-based AI agents can conduct structured interviews at scale, asking open-ended questions, following up with probing inquiries, and capturing nuanced qualitative feedback that written surveys typically miss \citep{wambsganss2020,kim2019}. These systems combine the scalability of automated surveys with the interactive richness of human interviews, potentially transforming how institutions gather and act on student input.

This research addresses a fundamental question: \emph{What are the characteristics, depth, and strategic value of feedback collected through voice-based conversational AI agents in higher education?} Specifically, we investigate whether conversational AI elicits detailed, honest, and actionable feedback that can serve as a foundation for instructional improvement and institutional learning. This question sits at the intersection of three theoretical domains: survey methodology and mode effects, human--computer interaction (particularly the CASA paradigm), and strategic management theory concerning organizational learning and dynamic capabilities.

The research gap is substantial. While extensive literature documents mode effects between telephone, face-to-face, and written surveys \citep{dillman2006,deleeuw1992,holbrook2003}, and emerging work explores conversational AI for data collection \citep{kim2019,sachdeva2024}, the characteristics and value of AI voice feedback in educational settings remain empirically underexplored. Moreover, the strategic implications of AI-augmented feedback systems for institutional learning capabilities have received limited attention. This study fills these gaps through a field study conducted in an authentic educational setting at a leading business school.

Our contributions are threefold. Theoretically, we extend CASA theory to AI-mediated educational feedback to test whether social responses to conversational AI enhance disclosure and elaboration in student evaluations. We also contribute to the emerging literature on AI-mediated data collection by characterizing voice-based AI as a distinct modality with unique affordances. Methodologically, we operationalize ``actionability'' through multiple coding dimensions---specificity, thematic richness, constructive orientation, and contextual depth---providing a framework for assessing feedback utility. Practically, we offer evidence-based guidance for institutions considering AI deployment in feedback systems, with implications for how educational institutions build dynamic capabilities through AI-augmented sensing mechanisms.

The remainder of this proposal proceeds as follows: Section~\ref{sec:theory} reviews the theoretical foundations of survey methodology, the CASA paradigm, and dynamic capabilities. Section~\ref{sec:design} details our research design, including the TUM pilot and MIT main study. Section~\ref{sec:contributions} outlines expected contributions, and Section~\ref{sec:status} discusses project status and timeline.

% ---------------------------------------------------------------------
\section{Theoretical Background}
\label{sec:theory}

\subsection{Survey Methodology and Mode Effects}

Survey mode---the medium through which questions are presented and answers collected---profoundly influences response quality, a phenomenon extensively documented in the survey methodology literature \citep{dillman2006,deleeuw1992,schwarz1991}. Mode effects arise from differences in cognitive processing, social context, and communication affordances across administration formats.

Cognitive and communicative mechanisms shape response behavior across modes. Self-administered questionnaires allow respondents more time for reflection, potentially increasing accuracy for factual recall \citep{schwarz1991}. Conversely, oral interviews impose time pressure, leading respondents to rely on more accessible information, though this can also reduce satisficing---the tendency to provide minimally acceptable answers \citep{chang2010}. \citet{holbrook2003}, in a landmark study, demonstrated systematic differences between telephone and face-to-face interviewing of national probability samples, establishing that mode shapes both the content and quality of responses.

Recent experimental evidence reveals complex tradeoffs between voice and text modalities. \citet{hohne2024} found that voice answers to open-ended questions contained significantly more words and covered more topics than text answers. \citet{lenzner2024} reported that oral answers to probing questions contained approximately three times more words and 0.3 more themes than written answers. However, \citet{schober2015} found in a large, randomized experiment ($N = 634$) that text interviews produced more differentiated responses and greater disclosure of socially undesirable information. \citet{kim2019}, in a highly cited study, directly compared chatbot and web survey data and found significant platform and conversational design effects on response quality.

For educational feedback, which involves both open-ended reflections on learning and potentially sensitive evaluations of instructors, these mode effects create competing predictions that our experiment is designed to adjudicate.

\subsection{The CASA Paradigm and Human--AI Interaction}

The Computers As Social Actors (CASA) paradigm posits that humans unconsciously apply social rules to computers, responding to them as social actors even when users recognize they are interacting with machines \citep{nass2000,nass1997}. This framework provides theoretical grounding for predicting how students will respond to AI voice interviewers.

Foundational work by \citet{moon2000}, published in the \emph{Journal of Consumer Research}, demonstrated that computers can elicit intimate self-disclosure through reciprocal exchange, establishing that the social dynamics of human conversation transfer to human--computer interaction. \citet{nass1997} showed that people exhibit politeness toward computers, providing more positive evaluations when the same computer asks about its own performance. \citet{lucas2014}, in a landmark study, demonstrated that virtual humans increase willingness to disclose, particularly for sensitive information, because respondents feel less judged by non-human interviewers.

More recent work has extended these findings. \citet{xiao2019} found that conversational agents that engage in self-disclosure elicit greater reciprocal disclosure from users. \citet{saffarizadeh2022} demonstrated that reciprocal self-disclosure by conversational agents increases both cognition-based and affect-based trustworthiness through anthropomorphism. Papneja and Yadav \citeyearpar{papneja2024} provide a comprehensive literature review and emergent framework for self-disclosure to conversational AI, identifying social presence, perceived anonymity, and conversational design as key moderators. Schroeder and Schroeder \citeyearpar{schroeder2018} further showed that the mode of interaction affects willingness to share personal information, with implications for trust in machine-mediated exchanges.

In the context of student feedback, CASA theory predicts that voice-based AI agents will trigger social norms of elaboration and reciprocity, potentially eliciting richer responses. However, social presence could also increase social desirability bias. Our study measures both response richness and honest disclosure to disentangle these competing effects.

\subsection{Dynamic Capabilities and Organizational Learning in Higher Education}

Beyond pedagogical improvements, AI-augmented feedback systems have strategic implications for how educational institutions build and sustain competitive advantage through organizational learning. Dynamic capabilities theory, particularly Teece's \citeyearpar{teece2007} sensing--seizing--reconfiguring framework, provides a lens for understanding these strategic dimensions. Dynamic capabilities refer to an organization's ability to integrate, build, and reconfigure internal and external competencies to address rapidly changing environments. In higher education, these capabilities are essential for adapting curricula, teaching methods, and student support systems in response to evolving student needs, labor market demands, and pedagogical innovations.

The ``sensing'' capability depends critically on the quality and timeliness of information flows from the environment. Traditional student feedback systems often provide delayed, sparse signals that limit institutional responsiveness. AI conversational interviewers offer the potential to enhance sensing capabilities in multiple ways. First, by eliciting richer, more detailed feedback, AI agents capture finer-grained signals about student experiences and emerging needs \citep{wambsganss2020,hohne2024}. Second, the scalability of AI systems enables more frequent feedback collection, potentially mid-semester or even weekly, allowing institutions to detect issues and opportunities earlier. Third, automated analysis of conversational data can identify patterns and themes across large student populations, surfacing insights that might be missed in manual review of written comments.

The ``seizing'' component involves translating feedback into concrete pedagogical adjustments---revising course materials, modifying teaching approaches, or providing targeted student support. AI-generated feedback that is more specific, contextual, and actionable directly supports this translation process by reducing the interpretive burden on instructors and administrators. The ``reconfiguring'' component involves building institutional processes and routines around continuous feedback loops. Deploying AI feedback systems represents a form of organizational reconfiguration, embedding new technological capabilities into teaching improvement workflows and potentially shifting institutional culture toward more data-driven, iterative course design.

The strategic value of AI feedback systems operates at multiple organizational levels. At the micro level, individual instructors gain richer data to refine their teaching. At the meso level, departments and programs can aggregate feedback to identify systemic issues, benchmark teaching quality, and allocate professional development resources. At the macro level, institutions build distinctive capabilities in educational innovation and responsiveness, potentially enhancing their reputations and competitive positioning. Our study focuses on micro-level outcomes, but the findings have clear implications for meso- and macro-level strategic decisions.

Integrating these theoretical streams yields a multi-layered model. CASA mechanisms explain how AI voice agents elicit different responses from students through social presence and reciprocity. The mode effects literature specifies the differences to expect in terms of richness, disclosure, and elaboration. Dynamic capabilities theory contextualizes why these differences matter strategically, linking micro-level feedback quality to institutional learning and adaptation capabilities. This integrated framework guides our research design and interpretation of findings.

% ---------------------------------------------------------------------
\section{Research Design and Methodology}
\label{sec:design}

\subsection{Technical Pilot Study at TUM}

To establish technical feasibility and refine the AI interviewer protocol, we conducted a pilot study at the Technical University of Munich (TUM) during the 2025--2026 academic year. The pilot was embedded in the course ``Value Creation with AI in Corporations: Lessons from Success and Failures,'' delivered in collaboration with Boston Consulting Group (BCG). This setting provided an ideal testbed: the course content centered on applied AI in organizational contexts, making students both receptive to AI-mediated interactions and able to provide informed feedback on the experience.

The pilot tested an ElevenLabs-based conversational AI system in a real educational setting, with three primary objectives: (1)~validate the technical infrastructure for conducting voice-based AI interviews, including telephony integration, conversation flow management, and data storage; (2)~assess student willingness to participate in AI-mediated feedback and identify potential barriers to adoption; and (3)~collect preliminary feedback data to inform refinement of interview protocols for the main study at MIT.

Students were invited to participate in telephone interviews with the AI agent about their course experiences. The agent introduced itself as a feedback interviewer, explained the purpose of the call, assured confidentiality, and proceeded through structured open-ended questions covering course content, teaching effectiveness, and suggestions for improvement. The system was programmed to ask follow-up probing questions based on student responses, such as ``Can you give me a specific example?'' or ``What would you recommend to improve that aspect?'' The pilot successfully demonstrated technical feasibility: the AI system reliably conducted interviews, managed conversational turns, and recorded responses without significant technical failures. Student participation rates were acceptable, and informal feedback indicated general comfort with the AI interviewer. Notably, the pilot generated a dataset of voice-based feedback that, while not yet formally analyzed, appears qualitatively richer than typical written survey comments. Based on pilot experiences, we refined the conversational opening to more clearly establish the AI's role, adjusted probing strategies to balance depth with interview length (targeting 5--7 minutes), and enhanced data security protocols to meet MIT's IRB requirements.

\subsection{Main Field Study at MIT Sloan}

The main study will be conducted in the ``Advanced Leadership Communication'' course (15.281) at MIT Sloan School of Management, with approximately 60 students. This setting offers a compelling context: the course focuses on communication skills, meaning students are likely attuned to the quality of interpersonal interactions, including those with AI agents, and the institutional context provides external validity for findings intended to generalize across elite higher education environments.

Unlike traditional experimental comparisons of survey modes, this study adopts an impact-focused design to examine the characteristics and value of voice-based AI feedback as a novel feedback modality in its own right. Rather than positioning AI voice feedback against conventional written instruments, we investigate what AI-mediated conversational feedback produces---its depth, richness, actionability, and the student experience of providing it. This design choice reflects our theoretical framing: we are less interested in which mode ``wins'' and more interested in understanding the unique affordances and outcomes of conversational AI for feedback collection, with implications for how institutions might integrate this capability into their feedback ecosystems.

Based on our theoretical framework, we advance three hypotheses:

\hyp{Hypothesis 1 (Elaboration through conversational affordance).}{Voice-based AI feedback interviews will yield responses of substantial linguistic depth, with students providing extended narratives, concrete examples, and multi-topic coverage that exceed typical benchmarks observed in the written course evaluation literature.}

\hyp{Hypothesis 2 (Social presence and disclosure).}{Students will perceive the AI interviewer as a socially present but non-judgmental interlocutor, leading to candid disclosure of both positive and critical feedback, including specific suggestions for improvement that students might withhold in instructor-adjacent evaluation settings.}

\hyp{Hypothesis 3 (Actionability and strategic utility).}{Feedback collected through voice-based AI interviews will demonstrate high actionability, characterized by specificity, constructive orientation, and contextual depth, making it directly useful for instructional improvement and institutional sensing processes.}

All students will be invited to participate in a structured telephone interview with the AI conversational agent. The AI conducts a 5--7-minute interview covering course content, teaching effectiveness, learning outcomes, and suggestions for improvement. Key questions include: ``What aspects of the course content did you find most valuable for your learning?'', ``How would you describe the teaching style and its effectiveness?'', and ``What recommendations would you offer to make this course more effective for future students?'' For each question, the AI asks follow-up probes if the initial response is brief, such as ``Could you elaborate on that?'' or ``What specifically made that valuable?'' All students will be informed that participation is voluntary, responses are confidential, and feedback will be used solely for course improvement and research purposes with appropriate IRB approval.

\subsection{Data Collection and Analysis}

Data collection will be managed through the Sona~8 platform, which provides a structured research participation system for recruitment, scheduling, and tracking. Primary data includes transcribed AI interviews, all de-identified for analysis. Metadata and engagement indicators include response length (word count per answer and total), completion rates, interview duration, and item nonresponse (questions declined or insufficiently addressed).

Approximately one week after feedback collection, participants will receive a follow-up survey assessing their experience: satisfaction with the feedback process, perceived ease of providing feedback, perceived confidentiality and psychological safety, and willingness to participate in similar feedback in the future. Additionally, the course instructor will review a sample of feedback and rate responses on actionability using a structured rubric, providing an external validity check.

Our analysis employs a convergent mixed-methods design. Two independent coders will analyze all feedback responses using a codebook operationalizing ``actionability'' across four dimensions: \emph{specificity} (references to specific course elements such as particular lectures, assignments, or readings versus generic comments), \emph{thematic richness} (number of distinct topics addressed per response), \emph{constructive orientation} (presence of concrete improvement suggestions versus purely evaluative judgments), and \emph{contextual depth} (inclusion of examples, explanations, or reasoning). Each dimension is coded on a 3-point scale (low, medium, high), and an overall actionability score is computed. Inter-rater reliability will be assessed using Cohen's kappa.

Quantitative analysis includes descriptive statistics on response characteristics (word count distributions, thematic coverage, completion patterns) and inferential tests of our hypotheses against established benchmarks from the written survey literature. We will apply automated sentiment analysis and linguistic complexity metrics (lexical diversity, syntactic complexity) to characterize the linguistic properties of AI-mediated feedback. Qualitative thematic analysis will identify emergent patterns---for instance, whether voice feedback includes more narrative or experiential content and how students construct improvement suggestions in conversational settings.

% ---------------------------------------------------------------------
\section{Expected Contributions}
\label{sec:contributions}

\paragraph{Theoretical contributions}
This research makes several contributions at the intersection of survey methodology, human--computer interaction, and strategic management. First, we extend CASA theory to educational feedback contexts, testing whether social responses to AI voice agents enhance disclosure and elaboration in student evaluations---a context involving evaluation, potential sensitivity, and institutional stakes that represents an underexplored boundary condition for CASA phenomena. Second, we advance the emerging literature on AI-mediated data collection by empirically characterizing voice-based conversational AI as a novel feedback modality with unique affordances \citep{sachdeva2024,wambsganss2022}. Third, our multi-dimensional coding framework for operationalizing feedback actionability, encompassing specificity, thematic richness, constructive orientation, and contextual depth, provides a replicable methodological contribution applicable across diverse organizational feedback contexts. Fourth, by integrating dynamic capabilities theory with CASA and mode effects, we bridge micro-level feedback interactions with macro-level institutional learning and adaptation capabilities.

\paragraph{Practical and strategic implications}
Beyond theoretical contributions, this research offers actionable insights for educational institutions and broader organizational contexts. Institutions exploring AI-mediated feedback currently lack rigorous empirical evidence on the characteristics and value of such approaches. Our findings will provide detailed data on what voice-based AI feedback produces---its depth, richness, and actionability---and under what conditions students engage meaningfully with conversational AI interviewers. The study identifies design principles for effective AI feedback systems, including optimal interview length, probing strategies, and conversational tone. For institutions seeking to strengthen dynamic capabilities, this research demonstrates how AI technologies can augment organizational sensing by capturing richer, more nuanced signals from stakeholders. The findings inform strategic decisions about deploying AI-augmented feedback systems at multiple institutional levels, from individual course improvement to program-wide quality assurance and institutional innovation strategy. These implications extend beyond higher education to employee surveys, customer feedback, and stakeholder engagement across organizational settings.

% ---------------------------------------------------------------------
\section{Preliminary Insights and Project Status}
\label{sec:status}

The TUM technical pilot has been completed, with data collected but not yet formally analyzed. The MIT main study is in the planning and design phase, with preliminary approval secured to experiment in the Spring 2026 semester. IRB approval is in process, with submission expected in March 2026.

Preliminary review of TUM pilot data suggests that AI voice interviews generated substantially longer and more detailed responses than typical written survey comments observed in prior semesters. Students appeared willing to engage in extended conversations with the AI agent, often providing narrative examples and detailed explanations that go well beyond the terse comments characteristic of written evaluations. These preliminary observations are encouraging and motivate the rigorous investigation planned for MIT.

The proposed timeline for the MIT study includes finalizing the experimental protocol and IRB approval (March 2026), student recruitment, randomization, and data collection (April 2026), follow-up satisfaction survey and data transcription (May 2026), qualitative coding and quantitative analysis (June--July 2026), and manuscript preparation (August 2026). We anticipate challenges, including potential differential participation rates between conditions and technical issues with voice recognition; we will implement robust error-handling and conduct sensitivity analyses.

This study represents the first phase of a broader research agenda on AI-augmented feedback systems in education. Future work will explore longitudinal effects of AI feedback on teaching improvement, comparative effectiveness across different course types and institutional contexts, and integration of AI feedback with institutional learning analytics platforms.

% ---------------------------------------------------------------------
\renewcommand{\refname}{References}


\begin{thebibliography}{99}
\setlength{\itemsep}{3pt}

\bibitem[Chang \& Krosnick(2010)]{chang2010}
Chang, L., \& Krosnick, J. A. (2010). Comparing oral interviewing with self-administered computerized questionnaires: An experiment. \textit{Public Opinion Quarterly, 74}(1), 154--167.

\bibitem[de Leeuw(1992)]{deleeuw1992}
de Leeuw, E. D. (1992). \textit{Data quality in mail, telephone, and face-to-face surveys} [Doctoral dissertation, Vrije Universiteit Amsterdam].

\bibitem[Dillman(2006)]{dillman2006}
Dillman, D. A. (2006). Why the choice of survey mode makes a difference. \textit{Public Health Reports, 121}(1), 11--13.

\bibitem[Heerwegh \& Loosveldt(2008)]{heerwegh2008}
Heerwegh, D., \& Loosveldt, G. (2008). Face-to-face versus web surveying in a high-internet-coverage population. \textit{Public Opinion Quarterly, 72}(5), 836--846.

\bibitem[Holbrook et~al.(2003)]{holbrook2003}
Holbrook, A., Green, M., \& Krosnick, J. (2003). Telephone versus face-to-face interviewing of national probability samples with long questionnaires. \textit{Public Opinion Quarterly, 67}(1), 79--125.

\bibitem[H\"ohne et~al.(2024)]{hohne2024}
H\"ohne, J. K., Schlosser, S., \& Krebs, D. (2024). Typing or speaking? Comparing text and voice answers to open questions in smartphone surveys. \textit{Social Science Computer Review, 42}(3), 688--707.

\bibitem[Kim et~al.(2019)]{kim2019}
Kim, S., Lee, J., \& Gweon, G. (2019). Comparing data from chatbot and web surveys: Effects of platform and conversational style. \textit{Proceedings of the 2019 CHI Conference on Human Factors in Computing Systems}, 1--12.

\bibitem[Lenzner et~al.(2024)]{lenzner2024}
Lenzner, T., H\"ohne, J. K., \& Neuert, C. (2024). Innovating web probing: Comparing written and oral answers to open-ended probing questions. \textit{Journal of Survey Statistics and Methodology, 12}(4), 891--912.

\bibitem[Lucas et~al.(2014)]{lucas2014}
Lucas, G. M., Gratch, J., King, A., \& Morency, L.-P. (2014). It's only a computer: Virtual humans increase willingness to disclose. \textit{Computers in Human Behavior, 37}, 94--100.

\bibitem[Moon(2000)]{moon2000}
Moon, Y. (2000). Intimate exchanges: Using computers to elicit self-disclosure from consumers. \textit{Journal of Consumer Research, 26}(4), 323--339.

\bibitem[Nass \& Moon(2000)]{nass2000}
Nass, C., \& Moon, Y. (2000). Machines and mindlessness: Social responses to computers. \textit{Journal of Social Issues, 56}(1), 81--103.

\bibitem[Nass et~al.(1997)]{nass1997}
Nass, C., Moon, Y., \& Green, N. (1997). Are people polite to computers? Responses to computer-based interviewing systems. \textit{Journal of Applied Social Psychology, 27}(5), 1093--1110.

\bibitem[Papneja \& Yadav(2024)]{papneja2024}
Papneja, H., \& Yadav, N. (2024). Self-disclosure to conversational AI: A literature review, emergent framework, and agenda for future research. \textit{Personal and Ubiquitous Computing, 28}, 453--477.

\bibitem[Sachdeva et~al.(2024)]{sachdeva2024}
Sachdeva, A., Kim, A., \& Dennis, A. R. (2024). Taking the chat out of chatbot? Collecting user reviews with chatbots and web forms. \textit{Journal of Management Information Systems, 41}(2), 432--461.

\bibitem[Saffarizadeh et~al.(2022)]{saffarizadeh2022}
Saffarizadeh, K., Boodraj, M., Alashoor, T., \& Kemerer, C. F. (2022). The impact of reciprocal self-disclosure on post-interaction trust in conversational agents. \textit{Journal of the Association for Information Systems, 23}(4), 839--869.

\bibitem[Schober et~al.(2015)]{schober2015}
Schober, M. F., Conrad, F. G., Antoun, C., Ehlen, P., Fail, S., Hupp, A. L., \ldots\ Zhang, C. (2015). Precision and disclosure in text and voice interviews on smartphones. \textit{PLOS ONE, 10}(6), e0128337.

\bibitem[Schroeder \& Schroeder(2018)]{schroeder2018}
Schroeder, J., \& Schroeder, M. (2018). Trusting in machines: How mode of interaction affects willingness to share personal information. \textit{Proceedings of the 51st Hawaii International Conference on System Sciences}.

\bibitem[Schwarz et~al.(1991)]{schwarz1991}
Schwarz, N., Strack, F., Hippler, H.-J., \& Bishop, G. (1991). The impact of administration mode on response effects in survey measurement. \textit{Applied Cognitive Psychology, 5}(3), 193--212.

\bibitem[Teece(2007)]{teece2007}
Teece, D. J. (2007). Explicating dynamic capabilities: The nature and microfoundations of (sustainable) enterprise performance. \textit{Strategic Management Journal, 28}(13), 1319--1350.

\bibitem[Wambsganss et~al.(2020)]{wambsganss2020}
Wambsganss, T., Winkler, R., S\"ollner, M., \& Leimeister, J. M. (2020). A conversational agent to improve response quality in course evaluations. \textit{Proceedings of the 2020 CHI Conference on Human Factors in Computing Systems}, 1--6.

\bibitem[Wambsganss et~al.(2022)]{wambsganss2022}
Wambsganss, T., Zierau, N., S\"ollner, M., K\"aser, T., Koedinger, K., \& Leimeister, J. M. (2022). Designing conversational evaluation tools. \textit{Proceedings of the ACM on Human-Computer Interaction, 6}(CSCW2), 1--37.

\bibitem[Xiao et~al.(2019)]{xiao2019}
Xiao, Z., Zhou, M. X., Liao, Q., Mark, G., \& Chi, E. (2019). Tell me about yourself: Using an AI-powered chatbot to conduct conversational surveys. \textit{ACM Transactions on Computer-Human Interaction, 27}(3), 1--37.

\end{thebibliography}
\end{document}